\documentstyle[11pt]{article}

\def\roughly#1{\mathrel{\raise.3ex\hbox{$#1$\kern-.75em%
\lower1ex\hbox{$\sim$}}}}

\def\lsim{\roughly<}
\def\gsim{\roughly>}
\def\be{\begin{eqnarray}}
\def\ee{\end{eqnarray}}
\def\susc{susceptibility}
\def\suscs{susceptibilities}
\def\Tr{{\rm Tr}}

\def\ben{\begin{enumerate}}
\def\een{\end{enumerate}}
\def\bitem{\begin{itemize}}
\def\eitem{\end{itemize}}

\def\thefootnote{\fnsymbol{footnote}}
\newcommand{\beq}{\begin{eqnarray}}
\newcommand{\eeq}{\end{eqnarray}}
\def\susc{susceptibility}
\def\Nc{$N_c$}
\def\la{\langle}
\def\ra{\rangle}
\def\bi{\begin{itemize}}
\def\ei{\end{itemize}}
\def\ie{{\it i.e}}
\def\eg{{\it e.g.}}
\def\etal{{\it et al}}
\def\del{\partial}
\def\L{{\cal L}}

\long\def\beginomit#1\endomit{}
\def\np{{Nucl. Phys.}}
\def\prl{Phys. Rev. Lett.}
\def\pr {Phys. Rev.}
\def\PR {Phys. Repts.}

\def\pl{Phys. Lett.}
\def\L{{\cal L}}

\begin{document}


\begin{titlepage}\begin{center}

\hfill{T94/100}
\hfill{nucl-th/9409003}

\vskip 0.6in
{\Large\bf CHIRAL SYMMETRY IN NUCLEAR PHYSICS}\footnote{Talk given at
{\it International Symposium on Medium Energy Physics},\  August 22-26, 1994,
Beijing, China.}
\vskip 1.2in
{\large  Mannque Rho}\\
\vskip 0.1in
{\large  \it Service de Physique Th\'{e}orique, CEA  Saclay}\\
{\large\it 91191 Gif-sur-Yvette Cedex, France}\\
\vskip .6in
\vskip .6in

{\bf ABSTRACT}\\ \vskip 0.1in
\begin{quotation}

\noindent The role of chiral symmetry in nuclear physics is summarized. The
topics treated are the chiral bag model for nucleon structure
resulting from large $N_c$ QCD, the pion cloud in chiral perturbation
theory for low-energy
electroweak nuclear response functions, ``swelled hadrons" in nuclear matter,
chiral symmetry restoration and pseudo-Goldstone meson condensation
in nuclear medium.

\end{quotation}
\end{center}\end{titlepage}
\renewcommand{\thefootnote}{\#\arabic{footnote}}
\subsection*{\it Introduction}
\indent

The quarks that enter into nuclei and hence figure in
nuclear physics are the u(p),
d(own) and possibly s(trange) quarks. These are called ``chiral quarks"
since they are very light at the scale of strong interactions. Both the u and
d quarks are less than 10 MeV, much less than the relevant scale which I will
identify with the vector meson (say, $\rho$) mass $\sim 1$ GeV. The s quark
is in the range of 130 to 180 MeV, so it is not quite light. In some sense,
it may be classified as ``heavy" as in the Skyrme model for hyperons but
the success with current algebras involving kaons also indicates that it can be
considered as chiral as the u and d are. In this talk, I will consider the
s quark in the same category although the results based on the lightness
assumption may not be very accurate.

If the quark masses are zero, the QCD Lagrangian has chiral symmetry
$SU(n_f)\times SU(n_f)$ where $n_f$ is the number of massless flavors.
However we know that this symmetry in the world we are living in, namely
at low temperature ($T$) and low density ($\rho$), is {\it spontaneously}
broken to $SU(n_f)_V$ giving rise to Goldstone bosons denoted $\pi$,
the pions for $n_f=2$, the pions,  kaons and $\eta$ for $n_f=3$.
In nature, the quark masses are not
strictly zero, so the chiral symmetry is explicitly broken by the masses,
and the bosons are pseudo-Godstones with small mass. Again in the u and d
sector, the pion is very light $\sim 140$ MeV but in the strange
sector, the kaon is not so light, $\sim 500$ MeV. Nonetheless we will
pretend that we have good chiral symmetry and rectify our mistakes
by adding symmetry breaking terms treated in a suitable way.

The theme of this talk is then that most of what happens in nuclei are
strongly controlled by this symmetry pattern. Indeed, it was argued many
years ago\cite{rhobrown81} that chiral symmetry should play a
crucial role in many nuclear processes, much more than
confinement and asymptotic freedom -- the other basic ingredients of QCD --
would. More recently, it has become clear that much of what we can understand
of the fundamental nucleon structure also follows from chiral symmetry and its
breaking. This was also anticipated sometime ago\cite{brtoday,br-comm88}.

In this talk, I would like to tell you more recent and quite exciting
new development in this line of work which suggests that the old idea, quite
vague at the beginning, is becoming a viable model of QCD in many-body nuclear
systems.
\subsection*{\it Nucleon: The Chiral Bag in QCD}
\indent

The chiral bag was formulated originally in a somewhat as hoc way based solely
on chiral symmetry but there is a striking indication\cite{mattis,manohar}
that it follows from a
more general argument based on large $N_c$ QCD where $N_c$ is the number of
colors. Let me discuss this as a model for nucleon structure.

When chiral symmetry is implemented to the bag model of the
hadrons\cite{br-comm88,chodos}, it was found necessary to introduce pion fields
outside of the bag of radius $R$ in which quarks are ``confined."
This is because otherwise the axial current cannot be conserved.
Furthermore, it was discovered\cite{rgb83} that to be consistent with
non-perturbative structure, the pion field takes the form of the
skyrmion configuration with a fractional baryon charge residing in the
pion cloud. This implied that the quarks are not strictly confined
in the sense of the MIT bag but various charges leak out. It became
clear that the bag radius is  not a physical variable. That physics should
not depend upon the size of the bag has been formulated as
a ``Cheshire Cat Principle" (CCP). In fact the CCP may be stated as a
gauge principle\cite{damgaardCCP} with the bag taken as a gauge fixing.
What this
meant was that the skyrmion is just a chiral bag whose radius is
``gauge-chosen" to be shrunk to a point.

The recent development\cite{mattis,manohar} is closer to the core of
QCD. In large $N_c$ QCD, meson-meson interactions become weak but
meson-baryon Yukawa interactions become strong going like $N_c^{1/2}$.
In this limit, the baryon is heavy and hence can be treated as a static
source localized at the origin. Other interactions, such as mass splittings
etc. are down by a factor of {\Nc}. Thus
we have to add to the usual current algebra
Lagrangian of $O(N_c)$ which I will denote $L_{ca}$
a term of the form\cite{manohar}
\be
\delta L=3g_A \delta(\vec{x}) X^{ia} A^{ia} (x)
\ee
where $X^{ia}$ is the baryon axial current in the large $N_c$ limit
and $A^{ia}$ is the pion axial current. It is found that summing an infinite
class of Feynman diagrams in the leading {\Nc} order corresponds to
solving coupled classical field equations given by the leading order
Lagrangian $L_{ca}+\delta L$.
This produces a baryon source coupled with a classical
meson cloud, with quantum corrections obtained by performing semiclassical
expansion around the classical meson background. This is precisely the
picture described by the chiral bag\cite{rgb83}.

There are two aspects of this result which are important for later purpose:
\bitem
\item It is conjectured -- and seems highly plausible --
that there is a line of UV fixed points in the large
{\Nc} renormalization group flow of the parameters of the
Lagrangian\cite{mattis}. The bag radius can be one of those parameters.
If correct, one may formulate CCP in terms of the ``fixed line."
\item The $m_\pi^3$ (or $m_q^{3/2}$) (where $m_\pi$ is the pion mass and $m_q$
the quark mass) non-analytic correction to the baryon mass that is found in the
classical solution is identical to a loop correction in chiral perturbation
theory\cite{manohar}. This makes a direct link between the chiral bag and
chiral perturbation theory ($\chi PT$) to a higher chiral order.
\eitem

\subsection*{\it Chiral Perturbation Theory for Nuclei}
\indent

As suggested above, the chiral bag links QCD to an effective Lagrangian
consisting of mesons and baryons. Here I will consider the u- and d-quark
system which provides a very nearly chirally symmetric situation. At low
energy, QCD then can be translated into chiral perturbation theory.
$\chi$PT is an expansion in derivatives and quark masses and this is known
to be very successful in describing $\pi\pi$ interactions\cite{leutwyler}.
In the presence of baryon fields, there is a complication because of the
baryon mass which is of order of the chiral scale $\Lambda_\chi\sim 1$ GeV.
In the pion sector, the expansion in $\del/\Lambda_\chi$ and
$\mu m_q/\Lambda_\chi^2$ (where $\mu$ is a scale parameter)
is all we need to do. However the derivative
$\del$ acting on the baryon field is of order of the baryon mass which is
not small compared with $\Lambda_\chi$. One way out is to use the heavy-baryon
formalism of Jenkins and Manohar\cite{JMHBF} which amounts to redefining the
baryon field by
\be
B\rightarrow e^{im_Bt}B.
\ee
This eliminates the mass $m_B$
when the derivative is taken, leaving small residual
momentum. This is an approximation that is clearly valid if the baryon is
very massive and propagates nearly on-shell. This amounts to an additional
expansion in power of $1/m_B$. The leading term in the expansion is just the
familiar static approximation.

As we saw above, the chiral bag structure anchored on large $N_c$ QCD and
$\chi PT$ with a heavy-baryon chiral Lagrangian describe the same physics
for the single baryon. We do not know how to use the chiral bag for nuclei
and nuclear matter but we are beginning to know how to do a systematic
$\chi PT$ for nuclear systems. One may say that doing $\chi PT$ is tantamount
to
doing QCD for nuclear physics.

The formalism has been applied to the calculation of nuclear forces
starting from a chiral Lagrangian that contains only the nucleon and pion
fields\cite{vankolck}. Going to one-loop order and using a momentum cut-off,
one can reasonably understand low-energy nucleon-nucleon interactions up to
$\sim 100$ MeV.

Here I would like to show the power of chiral Lagrangians
in describing exchange currents. One can formulate the same for electromagnetic
exchange currents but here I will confine the discussion
to axial charge transitions in nuclei.

Consider the $\beta$ transition between nucleus $A$ and nucleus $B$ of the type
\be
A(J^+)\leftrightarrow B(J^-),\ \ \ \ \Delta I=1
\ee
where  $I$ is isospin. This transition goes mainly through the time
component of the axial current, $J_{50}^i$. Warburton studied this
transition in the mass range $A=205$ -- $208$ with a
surprising result\cite{warburton}. Denote the matrix element of the single
particle operator associated with the axial charge by
$M_1$ and the experimentally extracted matrix element by $M^{exp}$.
Then the observation was that\cite{warburton}
\be
\frac{M^{exp}}{M_1}\sim 1.6\ -\ 2.\label{warburton}
\ee
Now the single-particle matrix elements are pretty carefully calculated,
so this big discrepancy cannot be in the nuclear wave functions. The conclusion
was that the operator must be deficient. How do we explain this?

The explanation was given many years ago\cite{kubodera} but without a good
understanding. It is only recently that a fully satisfactory answer
was found in terms of $\chi$ PT\cite{parketal}.

As shown by Weinberg in the case of nuclear forces\cite{weinberg}, if one
uses the heavy-baryon formalism, three-body and higher-body currents
do not contribute to next-to-leading order $\chi$PT (corresponding to
one loop) and hence we are left with only two-body corrections. Furthermore
to leading order, we have only the two diagrams of Fig.\ref{softpi} involving
one-pion exchange with {\it point-like} vertices. For the axial-charge
process,
\begin{figure}
\vskip 4cm
\caption{Soft-pion exchange two-body currents: the circled cross stands
for the current. For the electromagnetic current, both (a) and (b) contribute
while for the axial current, (b) does not contribute.}\label{softpi}
\end{figure}
it is even simpler since the diagram (b) vanishes by G-parity. Let us call this
soft-pion exchange contribution $M_2^{soft}$. Now next
to the leading order, there are many diagrams that can contribute in
general but for the axial charge transition, only a few graphs
survive. They are given in Fig.\ref{nextorder}. These are rather simple to
compute and are worked out by Park et al.\cite{parketal}.
\begin{figure}
\vskip 4cm
\caption{Next-to-leading order graphs that contribute to the axial
charge transition.}\label{nextorder}
\end{figure}
The result is that
\be
M^{th}=M_1 + M_2
\ee
with
\be
M_2= M_2^{soft} (1+\delta)
\ee
where the subscript represents the $n=1,2$-body operator.
It is found that to a good accuracy and almost independently of mass number,
\be
\delta\lsim 0.1.
\ee
Thus the loop corrections are
indeed quite small. Thus for the transition involved, the soft-pion
term dominates. A very similar situation holds for the magnetic dipole
process like the thermal np capture $n+p\rightarrow d+\gamma$  and seems to
hold
also for the process $e+d\rightarrow e+n+p$ even to a large momentum transfer.
This dominance of the soft-pion process in the cases considered was called
``chiral filter phenomenon." Calculation of the soft-pion term with realistic
wave functions\cite{ptk94} gives a large ratio
\be
M_2^{soft}/M_1\sim 0.6\ -\ 0.8
\ee
enough to explain the experimental value (\ref{warburton}).
This is the clearest indication that the pion cloud plays a crucial
role in nuclear processes.
\subsection*{\it ``Swelled" Hadrons in Medium}
\indent

The effective chiral Lagrangian I discussed so far is a Lagrangian
that results when the degrees of freedom lying
above the chiral scale $\Lambda_\chi$
are eliminated by ``mode integration." As one increases the matter density
or heats the matter,
the scale changes, so we can ask the following question: If a particle moves
in a background with a matter density $\rho$  and/or
temperature $T$, what is the effective Lagrangian applicable in this
background?
One possible approach is to take a theory defined at zero $T$ and zero
$\rho$ and compute what happens as $T$ or $\rho$ increases. This is the
approach
nuclear physicists have been using all along. However now we know that
the major problem with QCD is that the vacuum is very complicated and
we are not sure that by doing the standard approach we are actually describing
the vacuum correctly as $T$ or $\rho$ goes up.
Since the quark condensate $\la \bar{q}q\ra$ is a vacuum
property and it changes as one changes $T$ or $\rho$, it may be more profitable
to change the vacuum appropriate to the given $T$ or $\rho$ and build an
effective theory built on the changed vacuum.
This is the idea of Brown and Rho\cite{brscaling}
in introducing scaled parameters in the effective Lagrangian.

If the quarks are massless, then the QCD Lagrangian is scale-invariant
but quantum mechanically a scale is generated giving rise to the trace anomaly.
In the vacuum, we have in addition to the quark condensate $\la \bar{q}q\ra$
the gluon condensate $\la G_{\mu\nu}G^{\mu\nu}\ra$. We can associate
a scalar field $\chi$ to the $G^2$ field as $G^2\sim \chi^4$ and introduce
the $\chi$ field into the effective Lagrangian to account for the
conformal anomaly of QCD. The $\chi$ field can be decomposed roughly into
two components, one ``smooth" low frequency component and the other
``non-smooth" high-frequency component. The former can be associated
with 2-$\pi$, 4-$\pi$ etc. fluctuations and the latter with a scalar glueball.
For low-energy processes we are interested in, we can integrate out the
high-energy component and work with the low-energy one. How this is to be
done is explained in \cite{adamibrown}. The outcome of this operation is that
one can write the same form of the effective Lagrangian as in free space
with the parameters of the theory scaled as
\be
\frac{f_\pi^*}{f_\pi}\approx \frac{m_V^*}{m_V}\approx \frac{m_\sigma^*}
{m_\sigma}\approx\cdots\equiv \Phi (\rho) \label{brscaling1}
\ee
The nucleon effective mass scaled somewhat differently
\be
\frac{m_N^*}{m_N}\approx \sqrt{\frac{g_A^*}{g_A}}\frac{f_\pi^*}{f_\pi}.
\label{brscaling2}
\ee
In these equations the asterisk stands for in-medium quantity.
Now in the skyrmion model, at the mean-field level, $g_A^\star$ does not
scale, so the nucleon will also scale as (\ref{brscaling1}).
It turns out that the pion properties do not scale; the pion mass
remains unchanged in medium at low $T$ and $\rho$.
Thus if one of the ratios in (\ref{brscaling1}) is determined either by
theory or by experiment, then the scaling is completely defined. At
densities up to nuclear matter density, the scaling is roughly
\be
\Phi (\rho)&\approx& 1-a(\rho/\rho_0),\\
a&\approx& 0.15\ -\ 0.2\nonumber
\ee
where $\rho_0$ is the normal nuclear matter density.

Now given the Lagrangian with the scaled parameters, we can go on and do
loop corrections. One of the first things that one finds is that
the $g_A^\star$ gets reduced to $\sim 1$ in nuclear matter from 1.26 in
free space. So one would have to do the whole thing in a consistent way.
However the point is that most of the processes in nuclear physics are
dominated by tree-order diagrams and this means that the effective Lagrangian
with the scaled parameters should be predictive without further corrections.
Indeed this has been what has been found. In a recent paper, Brown, Buballa,
Li and Wambach\cite{BBLW} use this ``BR scaling" to explain simultaneously
the new deep
inelastic muon scattering experiment and Drell-Yan experiments.

A set of rather clear predictions has been made in this
theory.\cite{adamibrown,
elafmr}

\subsection*{\it Chiral Symmetry Restoration: The Georgi Vector Limit}
\indent

Naively pushed to the extreme, the BR scaling says that all the light-quark
hadron masses fall to zero as $T$ or $\rho$ reaches the chiral phase
transition point. Lattice gauge calculations are not yet in a position to
say anything about the high density regime but can tell us what happens
at high $T$: At $T=T_c\approx 140$ MeV, chiral symmetry gets restored.

Here is what we think happens in terms of the effective theory that I have
described\cite{br94}. At low $T$, we know that chiral $SU(2)\times SU(2)$
is broken down to $SU(2)_V$, so chiral symmetry is realized non-linearly
in the coset space $SU(2)\times SU(2)/SU(2)$. Now one can introduce a
hidden gauge symmetry\cite{bando} and
build an equivalent linear theory $[SU(2)\times SU(2)]_{global}\times
SU(2)_{local}$. Here we have, in addition to pions and constituent quarks or
nucleons, vector mesons
$\rho$ which are now gauge particles. (One can extend the symmetry to include
the $\omega$ meson.)

The $\rho$ mass in this theory is given by the well-known KSRF relation
\be
m_\rho=fg=\frac{1}{\sqrt{1+\kappa}} f_\pi g =
2\sqrt{1+\kappa} f_\pi g_{\rho\pi\pi}\label{KSRF}
\ee
where $\kappa$ is a temperature-dependent parameter which at
zero temperature takes the value $-1/2$, $g$ is the hidden gauge coupling
and $g_{\rho\pi\pi}$ is the $\rho\pi\pi$ coupling constant. Perturbative
calculation suggests that as $T$ approaches $T_c$, there is an ultraviolet
fixed point, $\kappa\rightarrow 0$.\cite{harada}
Georgi suggested that in this limit, the
$SU(2)\times SU(2)$ symmetry is restored but in a mode different from the usual
Wigner mode\cite{georgi}. Now the $SU(2)\times SU(2)$ symmetry here is
realized by the fact that the decay constants $f_\pi$
\be
\la 0 |A^i_\mu|\pi^j (q)\ra=i f_\pi q_\mu \delta^{ij}\label{goldstone}
\ee
and $f_S$
\be
\la 0|V_\mu^i| S^j (q)\ra= i f_S q_\mu \delta^{ij}
\ee
satisfy
\be
f_\pi=f_S.\label{mended}
\ee
In this scheme, there are two sets of Goldstone bosons, one set of
pseudoscalars (pions) and another set of scalars ($S$). For $g\neq 0$
and $f_\pi \neq 0$, the $\rho$ is massive, so the $S$ bosons make up
the longitudinal components of the massive $\rho$. Thus the symmetry
(\ref{mended}) in this situation resembles Weinberg's ``mended symmetry"
\cite{weinbergMS}.

The same perturbative calculation indicates that near $T_c$, the hidden
gauge coupling $g\rightarrow 0$. Georgi calls this the vector limit.
In this limit, the $\rho$ becomes massless, the $S$ bosons
are liberated and become degenerate with the pions. Since the gauge
coupling goes to zero, the gauge symmetry disappears and the local
symmetry gets swelled to global $SU(2)\times SU(2)$. Georgi proposed
this symmetry as possibly a large $N_c$ limit of QCD and that the deviation
from the vector limit be treated as perturbation. What we are proposing here
is that
this is a limit realized at the chiral phase transition temperature $T_c$.

Are there evidences that this is the correct scenario?

We suggest three observations\cite{br94}:
\ben
\item {\bf Quark number \susc}

Lattice gauge calculations\cite{gottlieb} show that both isoscalar
and isovector quark-number \suscs \ increase steeply from very small
value below $T_c$ to near free-quark value above $T_c$. This can be
understood simply if the vector mesons decouple from the quarks
across the transition temperature. Such a mechanism was suggested by
Kunihiro\cite{kunihiro} in the context of the NJL model. The Georgi
vector limit provides a concrete description of what might be happening.
The result is shown in Fig.\ref{gottliebchi} where the prediction is
compared with the lattice results.
\begin{figure}
\vskip 10cm
\caption{The isovector quark number \susc : the data are from the lattice
results
and the stars are the predictions based on the
Georgi vector limit.}\label{gottliebchi}
\end{figure}
\item {\bf Cool kaons in AGS heavy-ion collisions}

The recent preliminary
data\cite{stachel} on the $14.6$ GeV collision (experiment E-814)
\be
^{28}{\rm Si} + {\rm Pb}\rightarrow K^+ (K^-) + X
\ee
showed cool components with effective
temperature of 12 MeV for $K^+$ and 10 MeV for $K^-$. This cannot be
reproduced in the conventional scenarios employed in event generators
where one finds kaons of effective temperature $\sim 150$ MeV.

It is known experimentally that the freeze-out temperature of hadrons is
about 140 MeV which is about the same as the lattice result of the
chiral transition temperature $T_c$.
This suggests that the freeze-out for less strongly interacting
particles other than
the pion and the nucleon is at a temperature higher than $T_c$ and
that the pion and nucleon freeze out at about $T_c$. This means
that interactions in the interior of the fireball will be at temperature
greater than $T_c$.

Another point  is that the fireball must expand slowly.
The slow expansion results because the pressure in the region
for some distance above $T_c$ is very low \cite{kochbrown}, the
energy in the system going into decondensing gluons rather than giving
pressure.

Now what do we expect? As described later, $\chi PT$ to one-loop order
predicts that there are essentially three mechanisms that play an important
role in kaon-nuclear interactions:
(1) the $\omega$ meson exchange
giving rise to repulsion for $K^+ N$ interactions and attraction
for $K^- N$; (2) the ``sigma-term" attraction for both $K^\pm N$:
(3) the repulsive ``virtual pair term."
Roughly the vector-exchange gives the repulsion for $K^+$ (and an attraction
of somewhat more magnitude for $K^-$)
\be
V_{K^+ N}\cong \frac 13 V_{NN}\cong 90\ {\rm MeV}\,\frac{\rho}{\rho_0}
\label{repulsion}
\ee
where $\rho_0$ is nuclear matter density. This term is proportional to
the hidden gauge coupling $g^2$. One can estimate the scalar attraction
by the ``sigma term" (which is the same for both $K^\pm$)
\be
S_{K^+ N}\approx -\frac{\Sigma_{KN} \langle \bar{N}N\rangle}
{2 m_K f^2}\cong -45\ {\rm MeV}\,\frac{\rho_s}{\rho_0}
\label{attraction}
\ee
where $\rho_s$ is the scalar density and $\Sigma_{KN}$ is the $KN$ sigma
term.  The virtual pair term
(proportional to $\omega^2$ where $\omega$ is
the kaon frequency) -- which is
related to Pauli blocking -- removes, at zero temperature, about 60 \%
of the attraction (\ref{attraction}). At low temperature, the net effect is
therefore highly repulsive for $K^+ N$ interactions.

Now what happens as $T$ is equal or greater than $T_c$ is as follows.
First of all,
part of the virtual pair repulsion gets ``boiled" off as discussed in
\cite{BKR}. What is more
important, if the Georgi vector limit is relevant, then
the vector mesons decouple with $g\rightarrow 0$, killing off the repulsion
(\ref{repulsion}). As a consequence, the residual attraction from the
scalar exchange remains. This residual attraction combined with the high
freeze-out temperature is responsible for the cool components of $K^\pm$.

A recent calculation of Koch\cite{koch} which effectively models the
mechanism described here supports this
scenario. Given that the vector coupling is absent, one can see that
both $K^+$ and $K^-$ will have a similar cool component.
\item {\bf Instanton-molecule model for chiral phase transition}

As a final case, we mention a microscopic model that seems to
realize the Georgi vector symmetry at high temperature.

In a model where the chiral phase transition is described as a change
in the instanton liquid from a randomly distributed phase at low temperature
to an instanton-anti-instanton molecular phase above $T_c$, it has been
observed\cite{schafer} that the molecules get polarized in the time direction
and the interactions in the pion and the longitudinal vector channel
become identical. This leads to the degeneracy of the triplets of
$\pi$ and $\rho_\parallel$ which may be identified with the scalar $S$.
The interaction in the longitudinal vector-meson
channel becomes equally strong as attraction in
the scalar-pseudoscalar channel, while
transversely polarized vector mesons have no interaction.
If one assumes that the polarized molecules are the dominant agent for
interactions above $T_c$, then one can see that all coupling
constants in an NJL-type effective Lagrangian so generated could be
specified in terms of a {\it single} coupling constant, implying the swelling
of the symmetry in a manner closely paralleling the Georgi vector symmetry.
In this case, the restored
symmetry is $U(2)\times U(2)$ since the axial $U(1)_A$
is also supposed to be restored. Perturbative QCD effects are not expected
to modify this symmetry structure but it is not clear that no other
non-perturbative effects can enter to upset this. Nonetheless this is
a microscopic picture consistent with the Georgi vector symmetry.
\een
\subsection*{\it Kaon Condensation in ``Nuclear Stars"}
\indent

So far I have been concerned with the up and down quark hadrons.
I shall consider the case where the strange quark comes in directly.
Because of its non-negligible mass, low-order $\chi PT$ is probably
unreliable, so one would have to go at least to the next order.
This is what I will do here in connection with the condensation of kaons
in dense nuclear matter.

Kaon condensation is relevant to stellar collapse to compact stars,
customarily called ``neutron stars." We shall see that ``nuclear star"
would be a more appropriate nomenclature for this object.

$K^-$ condensation was originally predicted by Kaplan and Nelson
which was further supported by Politzer and Wise \cite{knpw}.
This was however done in tree order in $\chi PT$. Now tree-order
$\chi PT$ does not describe low-energy kaon-nucleon scattering
correctly and hence is a suspect when it comes to kaon condensation
which involves off-shell extrapolation.

What I want to report here is a next-to-next-to-leading order calculation
which describes correctly both kaon-nucleon scattering near threshold
{\it and} kaon-nuclear interactions ``seen" in
kaonic atoms\cite{LJMR,LBR,LBMR}.

Dimensional counting of Feynman diagrams using the effective Lagrangian
in heavy-baryon formalism allows one to classify each diagram for
the amplitude $KN\rightarrow KN$ with the integer
\be
\nu=1+2L+\sum _i (d_i +\frac{n_i}{2}-2)
\ee
where
$L$ is the number of loops, $d_i$ is the number of derivatives acting in
the $i$th vertex and $n_i$ the number of baryon lines entering the $i$th
vertex. Thus the dominant term is the one with no loop involving one
derivative of the form
\be
L_1 \sim \pm \frac{i}{f^2} K^\dagger \del_\mu K \bar{N} \gamma^\mu N,
\ \ \ {\rm for}\ \ \ K^\pm.\label{L1}
\ee
This term with $\nu=1$
is essentially the $\omega$-meson exchange between the $K$ and
$N$, the sign change representing the G-parity of the kaon involved.
This is the driving term for generating the $\Lambda (1405)$ state
in $K^- p$ scattering.
At the next order with $\nu=2$ involving no loops, there are two terms.
One is the ``sigma" term proportional to the quark mass matrix
\be
L_{sigma} \sim \frac{\Sigma_{KN}}{f^2} K^\dagger K \bar{N} N\label{Lsigma}
\ee
which is attractive for both $K^\pm$ channels. The other $\nu=2$
term is the two-derivative term at tree order,
\be
L_2 \sim (\del_\mu K)^2 \bar{N}N\label{L2}
\ee
which is repulsive for both $K^\pm$ in the s-wave channel.
Now loop terms bring in the factor $2L$ into $\nu$, so at one loop order, the
lowest contribution is given by $L=1$ with the Lagrangian (\ref{L1}).
To the same order, there are four-Fermi interactions contributing to the
energy density of the matter. It turns out that both one loop
and four-Fermi interaction terms are considerably smaller than the
tree-order contributions. In particular, the $\Lambda (1405)$ which plays
an essential role for threshold $K^- N$ interactions is irrelevant in
the condensation phenomenon. Consequently the net effect is then qualitatively
controlled by the three mechanisms (\ref{L1})-(\ref{L2}). This is just what
Kaplan and Nelson employed to predict the kaon condensation.

What happens physically is then as follows. In the collapse of a
massive star, as the matter is compressed, the electron chemical
potential goes up as the density piles up. Because of the attractive
interaction for $K^-$ which increases as density increases, the kaon mass
effectively falls. At some point, it becomes kinematically possible for
the electron to decay
\be
e^-\rightarrow K^- \nu_e.
\ee
If $K^-$ can condense, the decay can take place copiously. The charge
is neutralized by the protons, so when the kaons are condensed, the system
becomes more like a nuclear matter than neutron matter. It seems thus proper
to call a compact star resulting from this process a ``nuclear star."

The detailed calculation made by Lee {\etal}\cite{LBMR} predicts that
$K^-$ mesons will condense at
\be
\rho_c = (3\ - \ 4)\rho_0
\ee
where $\rho_0$ is the ordinary nuclear matter density $\approx \frac 12
m_\pi^3$. This falls in the range of density where interesting astrophysical
phenomena such as the formation of mini black holes as suggested by
Brown and Bethe\cite{brownbethe} can take place.

\subsection*{Acknowledgments}
\indent

What I discussed in this talk is recent development on some work
carried out since many years
in collaboration with Gerry Brown whom I would like to thank.
I have also benefited
from collaborations and/or discussions with many colleagues, in particular
Kuniharu Kubodera,
Chang-Hwan Lee, Dong-Pil Min, Maciek Nowak, Tae-Sun Park and Ismail Zahed.

\end{document}